\documentclass[11pt,twoside]{article}
\usepackage{asp2010}
\usepackage{epsf}
\usepackage{psfig}
\usepackage{lscape}

\begin{document}
\title{Bibliometric Evaluation of the Changing Finnish Astronomy}
\author{Eva Isaksson}
\affil{Helsinki University Library, Kumpula Campus Library, 
PO Box 64, 00014 University of Helsinki, Finland}

\begin{abstract}
This is a follow-up on the bibliometric evaluation of Finnish astronomy 
presented by the author at the LISA V conference in 2006. The data from 
the previous study are revisited to determine how a wider institutional 
base and mergers affect comparisons between research units.

\end{abstract}

Bibliometrics can be an useful tool for evaluating research done in
separate institutes working in the same field. The author has previously 
compared  four Finnish institutes in the field of astronomy for the period 
1995-2004 \citep{Isakssonp2007}.
Comparisons based on bibliometrics are, however, affected by many change factors. 
The bibliometric data itself can change -- e.g. affiliation data that was
sometimes sketchy or missing in the early 1990s is now more likely to be
included in research papers, as the administrative importance of evaluation 
and institutional comparisons is increasing.

\section*{Impact of Mergers}

What happens to institutional comparisons when the institutes change?
In Finland, the new University Act of 2009 imposed a lower limit to
the size of departments. This caused major changes in the structure
of Finnish universities and led to departmental mergers. As of 2010, 
the University of Helsinki Department of Astronomy has been 
merged with the Department of Physics. In Turku, the Tuorla Observatory 
merged with the Department of Physics in 2008 and became a separate unit in the new Department of Physics and Astronomy.

At the University of Oulu, the merger with the Department of Physics had already happened in 1998 
and therefore is out-of-step with our more current comparisons. In this study, we ask what 
sort of bibliometric approaches are useful when comparing astronomical 
research done before and after institutional mergers at the Helsinki and Turku 
universities. The short answer is: probably none. The long answer is 
that we can try to find ways to detect some trends.

\section*{Choosing \textit{A\&A}}

For the previous study, 918 refereed papers of the four biggest astronomy 
units at Finnish universities (Helsinki Observatory, Tuorla Observatory 
in Turku,  Oulu University and the Mets\"ahovi Observatory) were analyzed 
for 1995-2004. One of the findings was that 43\% of these papers were 
published in \textit{Astronomy \& Astrophysics}. 

For a second study, 622 \textit{A\&A} papers with a Finnish affiliation 
published between 1995 and 2009 and their citations were harvested from the Institute for Scientific Information.
Compared to the previous study that included complete 
publication data but required slow data mining, this was relatively easy to do. 
The use of just one journal guarantees that the affiliations are fairly 
consistent. Some physics research has been done by astronomy departments
while some astronomy has been done in physics departments. As all the papers 
in \textit{A\&A} can be considered astronomy papers, the question of whether 
one should draw a line between physics and astronomy papers can be ignored
for this study.

Can we foresee the impact of mergers by including \textit{A\&A} papers by 
the Helsinki Department of Physics and the rest of the Turku Department of
Physics and Astronomy with the Observatory papers in our analyses, so that we can compare 
astronomy research done by universities instead of by institutes?

\section*{Comparison by Papers}

From Figure 1, it is easy to see that both Turku and Helsinki benefit from
including the papers by astronomy and physics departments in our analyses, giving both
a stronger long-term lead compared to Oulu and Mets\"{a}hovi. 
It is worth noting that at Oulu, where the institutional 
merger happened in the late 1990s, giving Oulu an initial boost, the number 
of \textit{A\&A} papers has been decreasing recently.

\begin{figure}[!hb]
\plotone{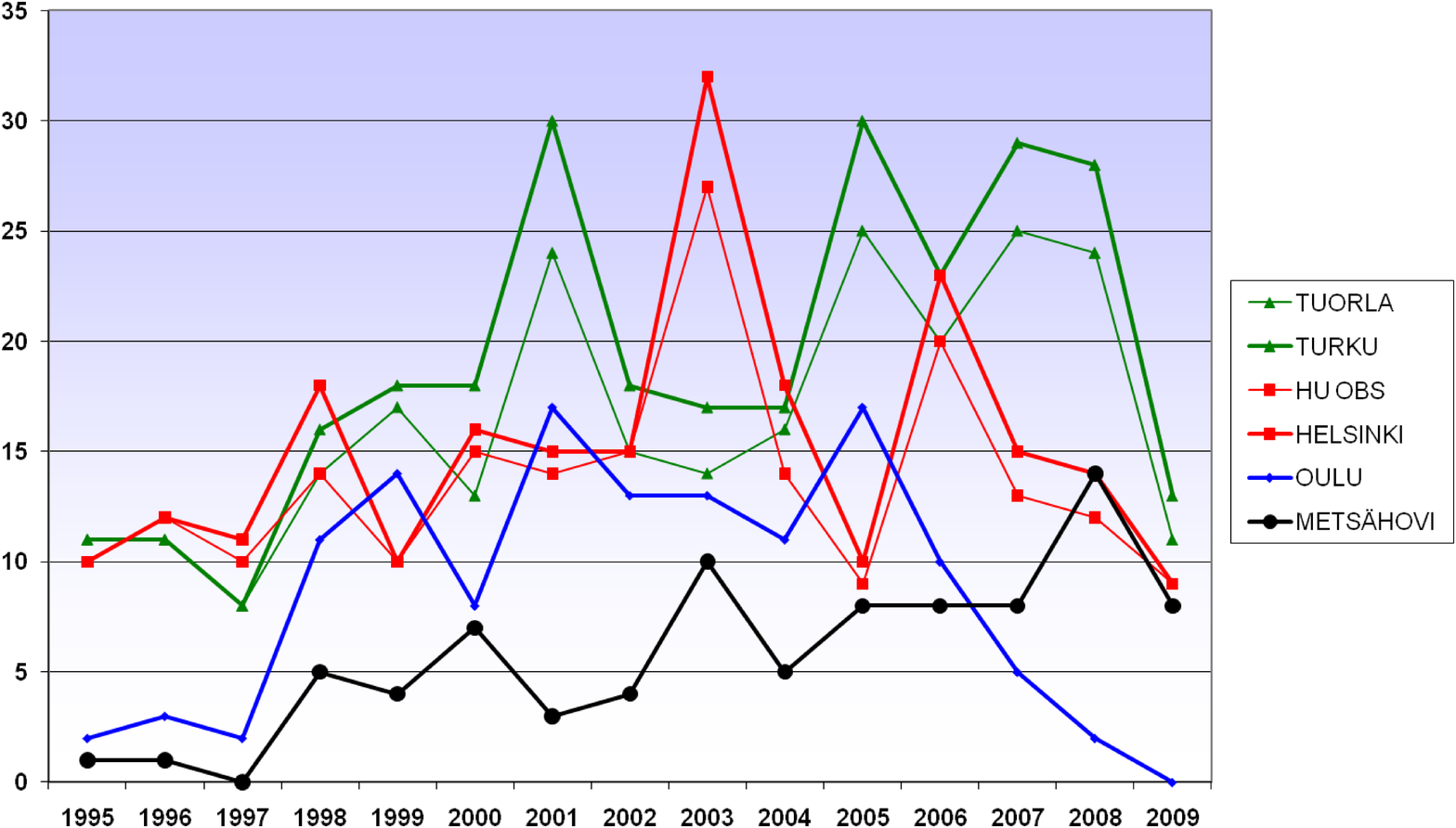}
\caption{Refereed Finnish \textit{A\&A}  papers per year according to ISI 
1995-2009. Tuorla refers to the Tuorla Observatory, HU Obs to the University 
of Helsinki Observatory. Turku and Helsinki are notably different after adding the \textit{A\&A} papers of their respective physics departments.}
\end{figure}

\section*{Comparison by Citations}

\begin{figure}[!ht]
\plottwo{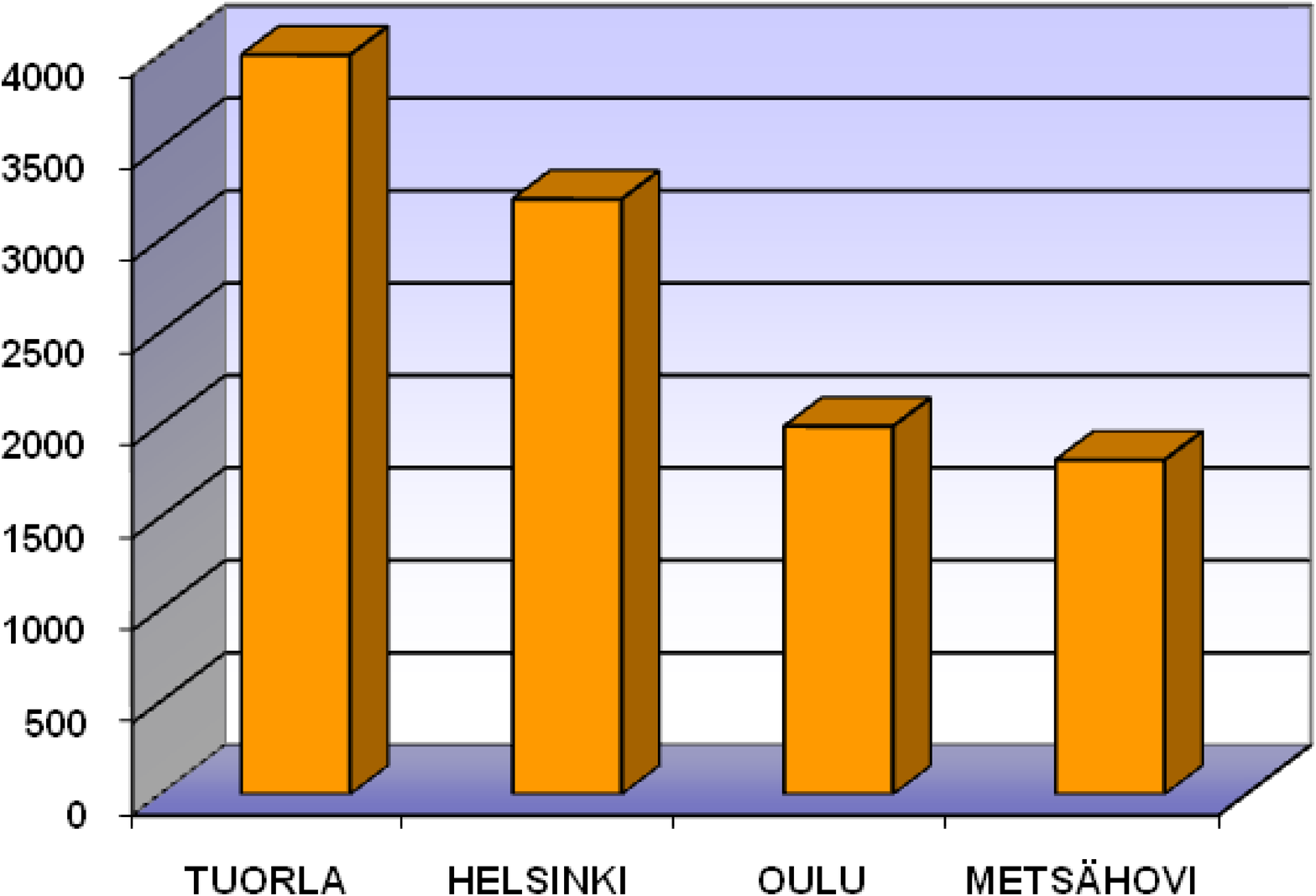}{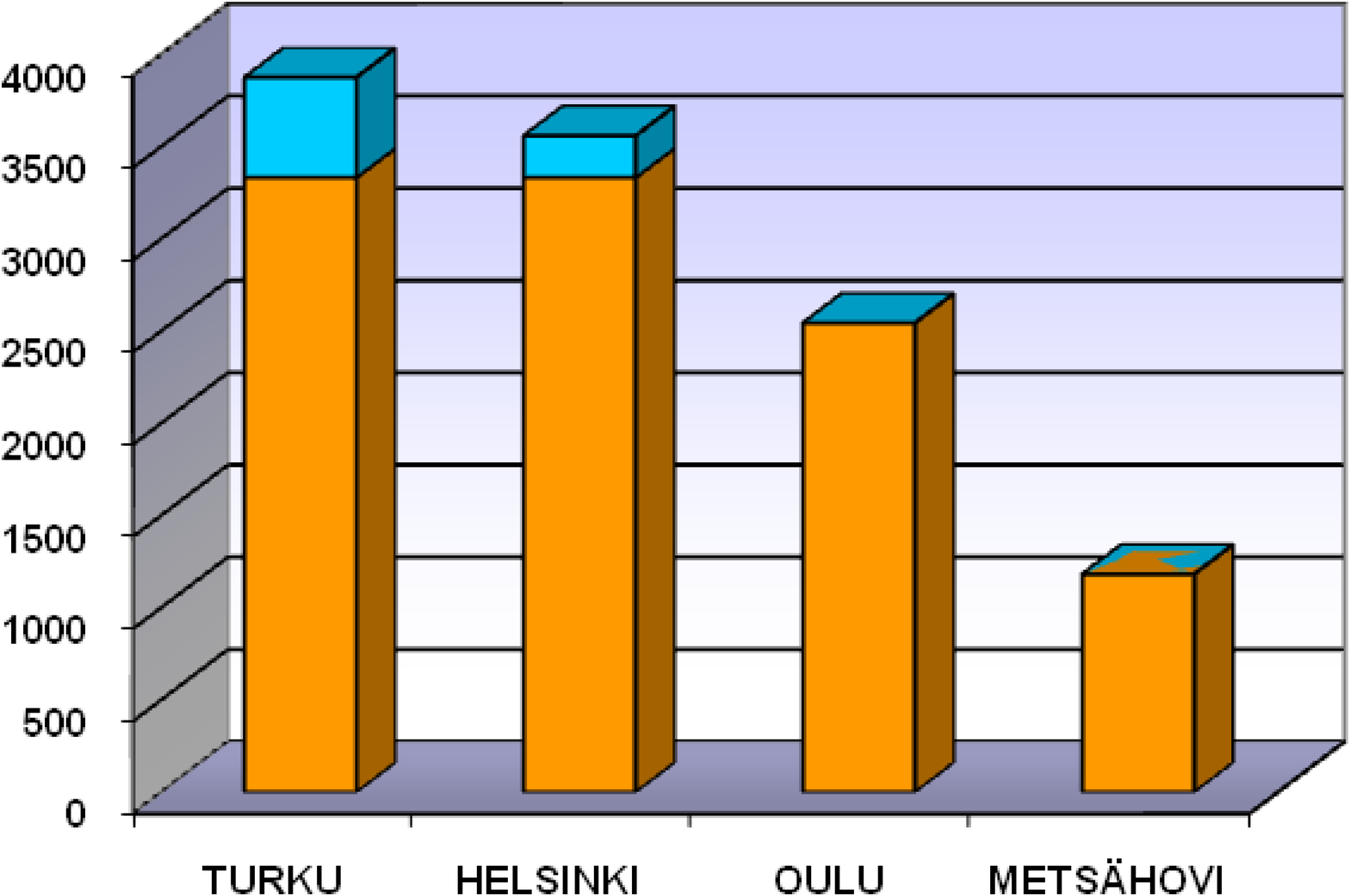}
\caption{Total citations 1995-2004 (left).
\textit{A\&A} citations 1995-2009 (right), showing the additional non-observatory 
papers.}
\end{figure}

Figure 2 shows us that the ranking of both the institutes and universities
is fairly consistent with the earlier, overall results. At the departmental 
level, both the Helsinki Observatory and the Tuorla Observatory received an 
equal number of citations. When we add physics papers, Turku gains quite visibly.  
One can ask whether it has benefited from its earlier institutional restructuring.

It also should be noted that the Mets\"ahovi \textit{A\&A} citations don't 
seem to quite match its publishing activity.

\section*{Conclusion}

In terms of \textit{A\&A} citations, both Tuorla and Helsinki have been 
consistently strong. But it is Tuorla that gains more from including the 
astronomy papers by people not working at the observatory. These results 
are interesting, but they are still based just on \textit{A\&A} papers. 
Including the papers from four additional journals would represent about 
70\% of all the papers and give more reliable results.

Is there really a trend? We cannot know until additional years have gone by 
to see what is happening, with new changes to blur the results.

\acknowledgements

The author has received the University of Helsinki Chancellor's travel grant for LISA VI. 
Thanks to Professor Karri Muinonen for asking the initial question : "how can we measure 
instutional change in bibliometric terms, " and to ``Pick U Up'' by Adam Lambert for 
cheering up the author during times of institutional turmoil.


\begin{thebibliography}{}
\bibitem[Isaksson(2007)]{Isakssonp2007}
Isaksson, E. 2007, in ASP Conf. Ser. Vol.
377, Library and Information Services in Astronomy V: Common
Challenges, Uncommon Solutions, ed. S. Richetts, C. Birdie, \& E.
Isaksson (San Francisco: ASP), 111


\end{thebibliography}
\end{document}